# Growth of metallic delafossite PdCoO$_2$ by molecular beam epitaxy

Matthew Brahlek[1*], Gaurab Rimal[2], Jong Mok Ok[1], Debangshu Mukherjee[3], Alessandro R. Mazza[1], Qiyang Lu[1], Ho Nyung Lee[1], T. Zac Ward[1], Raymond R. Unocic[3], Gyula Eres[1], Seongshik Oh[2*]

[1]Materials Science and Technology Division, Oak Ridge National Laboratory, Oak Ridge, TN, 37831, USA

[2]Department of Physics and Astronomy, Rutgers, the State University of New Jersey, Piscataway, NJ, 08854, USA

[3]Center for Nanophase Materials Sciences, Oak Ridge National Laboratory, Oak Ridge, TN, 37831, USA

Correspondence should be addressed to *brahlekm@ornl.gov and *ohsean@physics.rutgers.edu

**Abstract**: The Pd, and Pt based $AB$O$_2$ delafossites are a unique class of layered, triangular oxides with 2D electronic structure and a large conductivity that rivals the noble metals. Here, we report successful growth of the metallic delafossite PdCoO$_2$ by molecular beam epitaxy (MBE). The key challenge is controlling the oxidation of Pd in the MBE environment where phase-segregation is driven by the reduction of PdCoO$_2$ to cobalt oxide and metallic palladium. This is overcome by combining low temperature (300 °C) atomic layer-by-layer MBE growth in the presence of reactive atomic oxygen with a post-growth high-temperature anneal. Thickness dependence (5-265 nm) reveals that in the thin regime (<75 nm), the resistivity scales inversely with thickness, likely dominated by surface scattering; for thicker films the resistivity approaches the values reported for the best bulk-crystals at room temperature, but the low temperature resistivity is limited by structural twins. This work shows that the combination of MBE growth and a post-growth anneal provides a route to creating high quality films in this interesting family of layered, triangular oxides.

Confining 3-dimensional (3D) electron systems to 2D or 1D is a well-established route to observe and control quantum phenomena. Such effects arise naturally in a wide range of materials systems where the dimensionality of the crystal lattice naturally gives rise to 2D or 1D electronic structure. An interesting example is the unique $AB$O$_2$ delafossite oxides. The electronic structure is highly 2D due to natural $A$-$B$O$_2$ layering, where the $B$O$_2$ layer is composed of edge sharing $B$O$_6$ oxygen octahedra, and the $A$ layers are ionically bonded vertically above and below the oxygen atoms [1]. The delafossites come in two electronic motifs [2]. The first is the insulators, which have been extensively studied as $p$-type transparent conductors when degenerately doped [3,4]. Here, the $A$-site cations are, for example, Ag or Cu, and the $B$-site cations are typically either a transition metal or a trivalent $p$-block metal, for example, Al or Ga. The second class is the metallic compounds where $A$ = Pd or Pt, and the $B$-site cations are transition metals Co, Cr, or Rh (AgNiO$_2$, a delafossite polytype, is also metallic [5,6]). In the metallic phase the $A$-site cations are in a 1$^+$ valence state, which is unusual for elements like Pd and Pt. The low temperature conductivity reported in these metallic compounds are the highest reported for any oxide material, as high as $1.3\times10^8$ S/cm, and exhibit mean free paths in excess of 20 μm [7]. Further, the triangularly connected, highly localized states on the $B$O$_6$ octahedra offer an interesting platform for exploring effects of frustrated magnetism coupled with the itinerant electrons in the $A$-layer [8]. Together, these unique characteristics have combined to give rise to numerous exotic phenomena [6], including large interfacial Rashba-like spin splitting at the surface [9] as well as quantifying electron viscosity [9]. Yet, there is no report on growth of these materials by molecular beam epitaxy (MBE), and only one report by pulsed laser deposition (PLD) in the ultrathin limit [10]. Metallic delafossite films can be utilized for various electronic applications ranging from low-resistivity bottom electrodes for 2D materials to correlated transparent conductors [10,11]. Creation of well-controlled delafossite films will also enable answering many fundamental questions that cannot be readily addressed with bulk crystals such as the effects of strain on electronic and magnetic properties and dimensional confinement from 3D to 2D with reduced film thickness.

Here, we report the growth of the prototypical metallic delafossite PdCoO$_2$ by MBE on c-plane Al$_2$O$_3$ substrates (see Supplemental Materials for experimental details). The growth required overcoming the challenge of oxidizing palladium in the high temperature and ultrahigh vacuum environment typical of MBE. It was found that using atomic oxygen plasma at pressures up to $4\times10^{-6}$ Torr enabled $PdCoO_2$ to be grown only up to temperature of about 300 °C. Beyond this temperature, $PdCoO_2$ spontaneously decomposed by losing oxygen and forming Pd metal and CoO, as confirmed by surface sensitive reflection high-energy electron diffraction (RHEED) and x-ray photoemission spectroscopy (XPS), and bulk sensitive X-ray diffraction (XRD). As such, confining the growth to low temperature resulted in poor crystalline quality, thus necessitating a post-growth high-temperature anneal in an oxygen atmosphere. After annealing to $T_{Ann}$ = 700 °C in air and 800 °C in oxygen (as detailed in the Supplemental Materials, films were annealed in air or in oxygen and $T_{Ann}$, unless otherwise stated, refers to the air-annealing temperature), the crystalline quality improved substantially, the resistivity decreased and the surfaces became atomically flat. Above this temperature, however, the films were found to phase-segregate, likely driven by sublimation of volatile PdO. The balance dictated by the behavior of oxidation of Pd places a narrow window where high-quality single phase $PdCoO_2$ can be grown.

In order to find the optimal growth condition, various temperatures were used, and Fig. 1 shows examples of growth monitoring by RHEED at three different growth temperatures (300, 400, and 600 °C). The first attempts to grow $PdCoO_2$ were at a substrate temperature of 600 °C, which was motivated by two aspects: (1) For ternary oxides, typical growth temperatures around or above 600 °C are necessary to drive surface kinetics to enable crystallization, and (2) the reaction of Pd with oxygen to form volatile PdO [12] on the growing surface can enable adsorption controlled growth, for example, in $PbTiO_3$ [13,14] and ruthenate perovskites [15] (see also Ref. [16,17] and references therein for more examples); the possible formation of volatile PdO is highlighted by PLD growth of $PdCoO_2$ where synthesis of Pd deficient films necessitated alternating depositions from a $PdCoO_2$ target and a Pd-O target [10]. The real-time results from RHEED for deposition at 600 °C are shown in Fig. 1(a-c), which provides structural and morphological information. Starting with $Al_2O_3$ along the <100> crystal direction in Fig. 1(a), RHEED shows bright well-defined spots and Kikuchi lines, indicating flat morphology and high-structural quality. Under plasma oxygen at a pressure of $4\times10^{-6}$ Torr at 450 W, a monolayer (ML) of Co was initially deposited, followed by a ML of Pd, shown in Fig. 1(b-c), respectively. RHEED measurements showed that immediately the crystal quality degraded (loss of peak intensity), and 2D morphology became 3D (streaks gave way to spots), and the targeted phase was lost (additional peaks appeared off the main peaks). This was most apparent during the Pd deposition, which indicated metallic Pd clusters form on the surface rather than the oxide.

The formation of metallic Pd clusters suggests the growth temperature was too high, which motivated the following strategy. The rule-of-thumb for achieving a conformal epitaxial metal film requires the growth temperature to be around 3/8 of the melting point of the metal [18]. For Pd, this temperature is 400 °C, which would enable the formation of a conformal film of Pd, and, when deposited in an oxygen environment, likely enables formation of $PdCoO_2$. As such, it was found that a lower temperature of 400 °C combined with layer-by-layer deposition (Co-Pd-Co…Pd-Co), which followed the natural $PdCoO_2$ layering, was successful: on the other hand, codeposition failed to form the delafossite phase. To show the successful growth, RHEED images for the growth at 400 °C are shown in Fig. 1(d-f). In panel (d), at 3 ML the resulting film (*i*) is single phase (no extra spots off the main peaks), (*ii*) has a 2D morphology (streaky pattern), and (*iii*) is structurally consistent with $PdCoO_2$ (this point will be confirmed later). As the thickness was increased, however, additional crystal phases appeared, as can be seen in Fig. 1(e-f) for 8 ML and 18 ML; this is seen as the appearance of distinct half-order-like additional peaks, which resembles RHEED images previously reported for the growth of $Co_3O_4$ on $Al_2O_3$ [19] (see also Fig. S1 for corresponding XRD scan). This suggests that 400 °C is on the boundary of stability. It is interesting, however, that the film is initially stable in the thin limit but gradually becomes unstable as the thickness increases, suggesting that proximity to the substrate, thickness, or both affect the thermodynamic stability. By lowering the temperature to 300 °C, it was found with RHEED that single phase $PdCoO_2$ can be grown, as shown in Fig. 1(g) at 3 ML. As the thickness was increased, however, the morphology changed from 2D to 3D, but the phase remained that of $PdCoO_2$ (since no additional peaks appear), as can be seen by comparing Fig. 1(g-i) for thicknesses of 3 ML (g), 18 ML (h), and even up to 42 ML (i).

Figure 2(a) shows XRD $2\theta$-$\theta$ scans for PdCoO$_2$ films grown at 300 °C (labeled *as-grown*), as well as a sample that was annealed to 475 °C in vacuum. Other than the substrate peaks (marked by *), the lower black curve shows the 003$n$ set of peaks, where $n$ is an integer. This set of peaks is characteristic of the 3 ML periodicity of the unit cell of PdCoO$_2$ along the *c*-axis [1]. Furthermore, the lower, black-colored curves in Fig. 2(b) and (c) show XPS results for the as-grown film (see Supplemental Materials for experimental details), where Fig 2(b) and (c) show the results for Pd 3$d$ and Co 2$p$ states, respectively. The overall shape and binding energy agree with the results reported for bulk PdCoO$_2$ [20] with Pd$^{1+}$ and Co$^{3+}$. Both the XRD and XPS data agree with the interpretation of the RHEED data and confirm that the as-grown films at 300 °C are single phase PdCoO$_2$. These observations, however, indicate that the oxidation capabilities inherent to the MBE growth process (mainly pressure) limit the growth temperature and thus the structural quality of the PdCoO$_2$. Hence, a secondary post-growth anneal is likely to improve the quality.

To see this, we systematically explored annealing conditions in vacuum and *ex situ* in air and oxygen: we first discuss the results for vacuum annealing and later return to *ex situ* anneal. Shown in Fig. 2(b-c) are the temperature dependence of XPS spectra that were annealed in vacuum at 120, 360, 425, and 475 °C. With increasing annealing temperature, the peak positions of the Pd $3d_{3/2}$ and $3d_{5/2}$ shifted between 360 °C and 425 °C to lower binding energies. Similarly, Co $2p_{1/2}$ and $2p_{3/2}$ peaks shifted towards higher binding energies. This is consistent with the reduction of Pd$^{1+}$ to metallic Pd$^{0+}$, and the Co$^{3+}$ reducing to Co$^{2+}$ [20]. This is confirmed by XRD that was performed after the XPS anneal to 475 °C, the upper red curve shown in Fig. 2(a). The PdCoO$_2$ peaks were gone, and additional peaks appeared at 36.5°, 40.2°, 46.8°, and 77.7° in $2\theta$: the peaks at 36.5° and 77.7° are CoO, either the 101 and 202 peaks of wurtzite or the 111 or 222 peaks of rocksalt [21,22], and the peaks at 40.2° and 46.8° agreed, respectively, with the 111 and the 200 Pd peaks. This confirms that more oxidizing conditions are needed to improve the properties of PdCoO$_2$.

We now show the results for annealing PdCoO$_2$ films in air by starting with structural data and below discuss effects on transport data. The evolution of XRD $2\theta$-$\theta$ curves for as-grown PdCoO$_2$ and air-annealed to 700 °C and 800 °C are shown in Fig. 3(a). In comparing the as-grown and 700 °C curves, it can be seen that both are free from any secondary phases (for annealed films, it is noted that a peak emerges in some thick films at 61°, which may be related to formation of PdO). In going from 700 °C to 800 °C, an additional peak appears at around 39° in $2\theta$. This is most likely the 222 peak of $Co_3O_4$ [23]. The $Co_3O_4$ phase segregation is likely driven by the formation and evaporation of palladium oxide. The detailed structural evolution is shown in Fig. 3(b), where the intensity of the $PdCoO_2$ 006 and 0012 peaks and the 222 peak of $Co_3O_4$, normalized to the substrate 006 peak, are plotted versus annealing temperature. The intensity of the delafossite peaks are nominally constant until about 600 °C, and sharply increased in going from 600 °C to 800 °C by nearly an order of magnitude, indicating a large improvement to the structure of the film. However, by 800 °C the rise of the intensity of the $Co_3O_4$ peak shows that, although the structural quality of the $PdCoO_2$ continues to improve, it becomes unstable. When the film was annealed in pure oxygen instead, the film remained in single phase up to 800 °C (see Supplemental Materials).

High resolution XRD scans around the 006 and the 012 peaks of $PdCoO_2$, shown in Fig. 3(c-d), respectively, reveal two additional effects that occur during the annealing process. The first is the overall change of the in-plane and out-of-plane lattice parameters. In going from the as-grown film to the 700 °C and 800 °C, the 006 peak shifted towards smaller $2\theta$, indicating a shift toward larger out-of-plane lattice parameter (17.724→17.753 Å at 700 °C). As shown in Fig. 3(d), the in-plane lattice parameter showed the opposite trend, shifting toward larger $2\theta$, smaller lattice parameter (2.873→2.834 Å at 700 °C). The annealed values are closer to bulk $PdCoO_2$, indicating that either the films grow partially strained and annealing relaxes some of the residual strain, or defects, accumulated during the low temperature growth, are accommodated by lattice distortion and relieved through annealing. Strain due to the $Al_2O_3$ substrate with an effectively smaller in-plane lattice parameter of 2.763 Å (i.e. $4.785/\sqrt{3}$), would tend to contract the $PdCoO_2$ to an in-plane lattice parameter that is smaller than bulk (2.830 Å), which would thus cause an expansion of the out-of-plane lattice parameter to a value larger than bulk (17.837 Å). Since this is inconsistent with the experimental observation for the as grown films, the observed shift in lattice parameter is likely due to elimination of defects through annealing. Further, this also confirms that the films are fully relaxed with respect to the $Al_2O_3$ structure.

The second observation that can be made from the data shown in Fig. 3(c) is the absence of Laue oscillations in the as-grown films, their appearance at 700 °C, and finally their partial suppression at 800 °C. The origin of Laue oscillations is coherent scattering from the surface of a film and the interface between the film and substrate, which requires atomically smooth surfaces. As discussed previously, and shown in Fig. 1(i), the surfaces of the films grown at 300 °C become rough and 3D-like for thick films. This is consistent with the absence of Laue oscillations in the as-grown films. Annealing mobilizes the surface atoms, likely favoring the formation of the low-energy 001 plane of $PdCoO_2$. This further can be seen in RHEED measurements performed on ex situ annealed films, shown in Fig. 3(e) (the film was taken out of the MBE chamber, annealed, then put back into the MBE system for subsequent RHEED measurements). The comparison of the as-grown film (top of the panel) with the post-annealed (bottom), shows that the surface flattened and the crystal quality increased (indicated by much sharper diffraction peaks and appearance of Kikuchi lines); it is noted that the post annealed films exhibit a weak second-order reconstruction. For the film annealed to 800 °C the Laue oscillations were suppressed and became barely visible. This indicates surface roughening, which correlates with the XRD data shown in Fig. 3(a), where $Co_3O_4$ appeared at 800 °C. The two factors are likely connected and provide interesting insight into the synthesis of $PdCoO_2$. The $Co_3O_4$ formation was likely formed by out-diffusion of palladium oxide, which may be predominately confined to the surface, and as such, gives rise to roughening of the surfaces as $Co_3O_4$ nucleation likely starts there. In comparison to synthesis of bulk $PdCoO_2$, which utilizes air annealing up to 950 °C [1,2,24], the observed formation of $Co_3O_4$ at 800 °C in thin films further suggests that it is confined to the surface region (this conjecture is also supported by transport measurements, discussed next). More precisely, if this was dictated by equilibrium thermodynamics, bulk crystals would show similar impurities, and it is possible that bulk $PdCoO_2$ crystals likely have intrinsically Co-rich surfaces.

The residual resistivity for temperature near absolute zero is a very sensitive probe of crystal quality for metals as it only depends on the density of static disorder (crystal defects). Figure 4(a) shows temperature dependent resistivity for the same 25 nm-thick $PdCoO_2$ used for the XRD measurements in Fig. 3. The overall trend is a decrease in resistivity with decreasing temperature, reaching a finite resistivity as the temperature goes towards zero, which is typical of a metal. The overall trend with annealing is that the resistivity drops with increasing annealing temperature. The room temperature resistivity decreases with increasing annealing temperature, going from around 98.0 µΩcm for the as-grown films down to about 20.0 µΩcm for the films annealed to 700 °C in air. For films annealed in oxygen at 800 °C, the room temperature resistivity drops further to 4.7 µΩcm (see Supplemental Materials), which is comparable to the bulk crystal values, 2-8 µΩcm [6,7,24–27]. With annealing, the low temperature value changed from 90.0 µΩcm for the as-grown film down to 13.0 µΩcm with 700 °C air annealing and to 1.1 µΩcm with 800 °C oxygen annealing.

To expound on the evolution with increasing annealing temperature in air, the residual resistivity taken at $T = 7$ K, $\rho_{7K}$, is plotted versus annealing temperature $T_{Ann}$ in Fig. 4(b). With increasing annealing temperature, $\rho_{7K}$ does not change in going from as-grown to 300 °C, but beyond this it monotonically drops until 800 °C. Further, for the as-grown films it can be seen that $\rho_{7K}$ does not coincide with the minimum resistivity, as the minimum occurs at finite temperature $T_{Min}$, below which there was a very shallow increase in the resistivity with decreasing temperature: the values of $T_{Min}$ are plotted versus $T_{Ann}$ in Fig. 4(c). The resistive minimum originates from, most likely, disorder-induced localization effects. This is consistent with the behavior of $T_{Min}$ shown in Fig. 4(b), where $T_{Min}$ is constant at about 40 K to 300-400 °C, and monotonically decreases towards zero up to 800 °C. Overall the dependence of $\rho_{7K}$ and $T_{Min}$ on $T_{Ann}$ parallel each other, but differs with respect to the structural evolution shown in Fig. 3(b); more precisely, the structure is nominally constant up to about 600 °C, above which the diffraction peaks increase as the crystal structure improves. The lower onset for the drop in $\rho_{7K}$ and $T_{Min}$ with increasing $T_{Ann}$ suggests that defects begin to reduce right above the growth temperature (300 °C). The difference between transport and XRD with annealing stems from the fact that transport and XRD are not sensitive to defects in the same way. For example, transport shows that defects began to be eliminated above 300 °C, but this is only evident in XRD for temperatures above 600 °C, pointing to either multiple types of defects with separate temperature dependency or that transport is more sensitive to defects; this interesting point requires a future systematic study of defect structures in PdCoO$_2$ thin films. Lastly, returning to the issue of Co$_3$O$_4$ segregation in the 800 °C sample, the improvement in resistivity in going from 700 °C to 800 °C shows that the $Co_3O_4$ segregation is not the limiting defect for the transport properties of these films, suggesting that they form highly localized islands, possibly confined to the surface: note also that with oxygen annealing, the film remains in single phase up to 800 °C (see Supplemental Materials).

To understand the growth mode and defect structures in annealed films, thickness-dependent transport measurements are shown in Fig. 5 (a-b), where $\rho_{7K}$ and RRR are plotted versus thickness, respectively. These data show that between 5-75 nm there is a strong thickness dependence, implying that the dominant scattering mechanism is from the surface. Above a thickness of about 100 nm, the resistivity saturated to a value of about 0.25 µΩcm, which is the lowest among any metallic oxide thin films but still well-above the best values attainable in bulk $PdCoO_2$ crystals (0.0075 µΩcm) [7]. This is further reflected by the RRR value which saturated to a maximum value of 16, whereas bulk crystals can reach as high as ~400 [7,26].

Saturation of the resistivity in the thick limit indicates that bulk defects, instead of surface scattering, are the limiting factor of the transport at low temperature in thick films. The most likely factor is epitaxial twins. These can be seen in the azimuthal ($\phi$) scan about the {012} family of XRD peaks, which is shown in Fig. 5(c) for the $PdCoO_2$ film annealed to 700 °C. The $Al_2O_3$ substrate shows only 3 peaks, reflecting the 3-fold symmetry. In contrast, the $PdCoO_2$ film shows 6 peaks, indicating that there are two structural twins, labeled $T_1$ and $T_2$. The 012 $T_1$ and the 012 $T_2$ peaks of $PdCoO_2$ peaks appear at $\phi= \pm30°$ relative to the 012 peak of the $Al_2O_3$ structure. These twin domains can be explicitly seen in high-angle annular dark-field scanning transmission electron microscopy (HAADF STEM) images taken on an annealed film along the <100> crystal direction, shown in Fig. 5(d-e) (see Supplemental Materials for experimental details). In the wide scan, shown in Fig. 5(d), the $Al_2O_3$ substrate is the darker region and the $PdCoO_2$ film is the lighter upper portion. Here, the brightest atoms are the heavier Pd, and the lighter Co are the more diffuse atoms. In this image the high crystalline quality of these films is apparent, as the layered $PdCoO_2$ structure is clearly visible. Moreover, domain boundaries are clearly visible both vertically and horizontally relative to the $PdCoO_2$ layers. In the zoomed-in image in Fig. 5(e) (taken from the box in Fig. 5(d)) these boundaries are highlighted by yellow arrows. These domains are clearly identified as $T_1$, and $T_2$, by overlaying the crystal structure of PdCoO$_2$ projected along the <100> direction, and <-100> direction on top of the STEM data, where there is excellent agreement on the atom positions. The origin of the twins is either due to equivalent ways for both $T_1$ and $T_2$ to epitaxially match Al$_2$O$_3$, as shown in the Fig. S5 in the Supplemental Materials, the formation of stacking faults (see for example refs. [28,29]), or a combination of the two. Such twins are also observed in previous PLD grown films [10].

To conclude, we have shown that, although challenging, there is a route to growing PdCoO$_2$ epitaxial films by MBE on Al$_2$O$_3$. This required overcoming the primary challenge of oxidizing Pd. It was found that this limited the *in situ* growth temperature to about 300 °C, which was insufficient to drive diffusion, and thus, necessitates an *ex situ* post-growth anneal to high temperatures to create films with high crystalline quality. These films exhibit improved transport properties compared with the previous PLD grown films [10]. Specifically, the room temperature resistivity of the annealed films approaches that of the best single crystals, suggesting comparable crystalline quality, as confirmed by the STEM data. However, the transport properties at low temperature were likely limited by the formation of structural twins, which originates from the epitaxial relationship between the film and the substrate: there are two equivalent motifs to orient PdCoO$_2$ on Al$_2$O$_3$. In the context of future research directions, this work points out several key factors. (1) Growing the full class of metallic delafossites, including the Pt compounds and AgNiO$_2$, is likely more challenging than the Pd-based material shown here, since both Pt and Ag are more difficult to oxidize than Pd. But as the current work has shown, it may be possible with a low temperature growth followed by a post anneal in high oxygen pressure. Moreover, since Pt-oxides and Ag-oxides are non-volatile, unlike PdO, higher oxygen pressures and higher temperatures may enable formation of high quality films of these delafossites. (2) Improving crystal quality of PdCoO$_2$ may require moving away from Al$_2$O$_3$ as a substrate. Although a delafossite-based buffer may seem viable, such domains would always be present as they result from the structural mismatch between delafossite and Al$_2$O$_3$; thus, realizing single domain PdCoO$_2$, and other metallic delafossites, likely necessitates a development of delafossite single crystal substrates. (3) Due to the rich surface physics in these materials [9,30], there are many open questions regarding structural and electronic effects of surface termination and layering sequence. Altogether, the successful synthesis of high-quality metallic delafossite films as demonstrated here will enable a range of fundamental studies that cannot be accessed with bulk crystals in this rich family of layered, triangular oxides.

## Acknowledgements

This work was supported by the U.S. Department of Energy (DOE), Office of Science, Basic Energy Sciences (BES), Materials Sciences and Engineering Division (MBE synthesis, physical characterization) and BES Computational Materials Sciences Program (electronic structure characterization). J.M.O was supported by the Laboratory Directed Research and Development Program (x-ray diffraction) of Oak Ridge National Laboratory, managed by UT-Battelle, LLC, for the U.S. DOE. The electron microscopy work was performed as a user project at the Center for Nanophase Materials Sciences, a U.S. Department of Energy Office of Science User Facility (D.M. and R.R.U.). G.R. and S.O. are supported by the Gordon and Betty Moore Foundation's EPiQS Initiative (GBMF4418). M.B. would like to acknowledge QuantumEmX award from ICAM and the Gordon and Betty Moore Foundation through Grant GBMF5305 for travel support. We would like to thank Scott Chambers for assistance in interpreting XPS spectra.

## References

[1]     R. D. Shannon, C. T. Prewitt, and D. B. Rogers, Inorg. Chem. **10**, 719 (1971).

[2]     R. D. Shannon, D. B. Rogers, C. T. Prewitt, and J. L. Gillson, Inorg. Chem. **10**, 723 (1971).

[3]     A. Stadler, Materials (Basel). **5**, 661 (2012).

[4]     H. Kawazoe, M. Yasukawa, H. Hyodo, M. Kurita, H. Yanagi, and H. Hosono, Nature **389**, 939 (1997).

[5]     Y. J. Shin, J. P. Doumerc, P. Dordor, C. Delmas, M. Pouchard, and P. Hagenmuller, J. Solid State Chem. **107**, 303 (1993).

[6]     A. P. Mackenzie, Reports Prog. Phys. **80**, 032501 (2017).

[7]   C. W. Hicks, A. S. Gibbs, A. P. Mackenzie, H. Takatsu, Y. Maeno, and E. A. Yelland, Phys. Rev. Lett. **109**, 116401 (2012).

[8]   L. Balents, Nature **464**, 199 (2010).

[9]   V. Sunko, H. Rosner, P. Kushwaha, S. Khim, F. Mazzola, L. Bawden, O. J. Clark, J. M. Riley, D. Kasinathan, M. W. Haverkort, T. K. Kim, M. Hoesch, J. Fujii, I. Vobornik, A. P. Mackenzie, and P. D. C. King, Nature **549**, 492 (2017).

[10]  T. Harada, K. Fujiwara, and A. Tsukazaki, APL Mater. **6**, 046107 (2018).

[11]  L. Zhang, Y. Zhou, L. Guo, W. Zhao, A. Barnes, H.-T. Zhang, C. Eaton, Y. Zheng, M. Brahlek, H. F. Haneef, N. J. Podraza, M. H. W. Chan, V. Gopalan, K. M. Rabe, and R. Engel-Herbert, Nat. Mater. **15**, 204 (2016).

[12]  C. R. Abernathy, S. J. Pearton, F. Ren, W. S. Hobson, T. R. Fullowan, A. Katz, A. S. Jordan, and J. Kovalchick, J. Cryst. Growth **105**, 375 (1990).

[13]  C. D. Theis, J. Yeh, D. G. Schlom, M. E. Hawley, and G. W. Brown, Thin Solid Films **325**, 107 (1998).

[14]  E. H. Smith, J. F. Ihlefeld, C. A. Heikes, H. Paik, Y. Nie, C. Adamo, T. Heeg, Z.-K. Liu, and D. G. Schlom, Phys. Rev. Mater. **1**, 023403 (2017).

[15]  H. P. Nair, Y. Liu, J. P. Ruf, N. J. Schreiber, S.-L. Shang, D. J. Baek, B. H. Goodge, L. F. Kourkoutis, Z.-K. Liu, K. M. Shen, and D. G. Schlom, APL Mater. **6**, 046101 (2018).

[16]  D. G. Schlom, APL Mater. **3**, 062403 (2015).

[17]  M. Brahlek, A. Sen Gupta, J. Lapano, J. Roth, H.-T. Zhang, L. Zhang, R. Haislmaier, and R. Engel-Herbert, Adv. Funct. Mater. **28**, 1702772 (2018).

[18]  C. P. Flynn, J. Phys. F Met. Phys. **18**, L195 (1988).

[19]  C. A. F. Vaz, V. E. Henrich, C. H. Ahn, and E. I. Altman, J. Cryst. Growth **311**, 2648 (2009).

[20]  H.-J. Noh, J. Jeong, J. Jeong, E.-J. Cho, S. B. Kim, K. Kim, B. I. Min, and H.-D. Kim, Phys. Rev. Lett. **102**, 256404 (2009).

[21]  R. W. Grimes and K. P. D. Lagerlof, J. Am. Ceram. Soc. **74**, 270 (1991).

[22]  M. J. Redmanand E. G. Steward, Nature **193**, 867 (1962).

[23]  W. L. Smith, A. D. Hobson, and IUCr, Acta Crystallogr. Sect. B Struct. Crystallogr. Cryst. Chem. **29**, 362 (1973).

[24]  R. D. Shannon, D. B. Rogers, and C. T. Prewitt, Inorg. Chem. **10**, 713 (1971).

[25]  M. Tanaka, M. Hasegawa, and H. Takei, Czechoslov. J. Phys. **46**, 2109 (1996).

[26]  H. Takatsu, S. Yonezawa, S. Mouri, S. Nakatsuji, K. Tanaka, and Y. Maeno, J. Phys. Soc. Japan **76**, 104701 (2007).

[27]  H. Takatsu, J. J. Ishikawa, S. Yonezawa, H. Yoshino, T. Shishidou, T. Oguchi, K. Murata, and Y. Maeno, Phys. Rev. Lett. **111**, 056601 (2013).

[28]  R. Ramlau, R. Schneider, J. H. Roudebush, and R. J. Cava, Microsc. Microanal. **20**, 950 (2014).

[29]  J. H. Roudebush, N. H. Andersen, R. Ramlau, V. O. Garlea, R. Toft-Petersen, P. Norby, R. Schneider, J. N. Hay, and R. J. Cava, Inorg. Chem. **52**, 6083 (2013).

[30]  F. Mazzola, V. Sunko, S. Khim, H. Rosner, P. Kushwaha, O. J. Clark, L. Bawden, I. Marković, T. K. Kim, M. Hoesch, A. P. Mackenzie, and P. D. C. King, Proc. Natl. Acad. Sci. U. S. A. **115**, 12956 (2018).

Figure 1

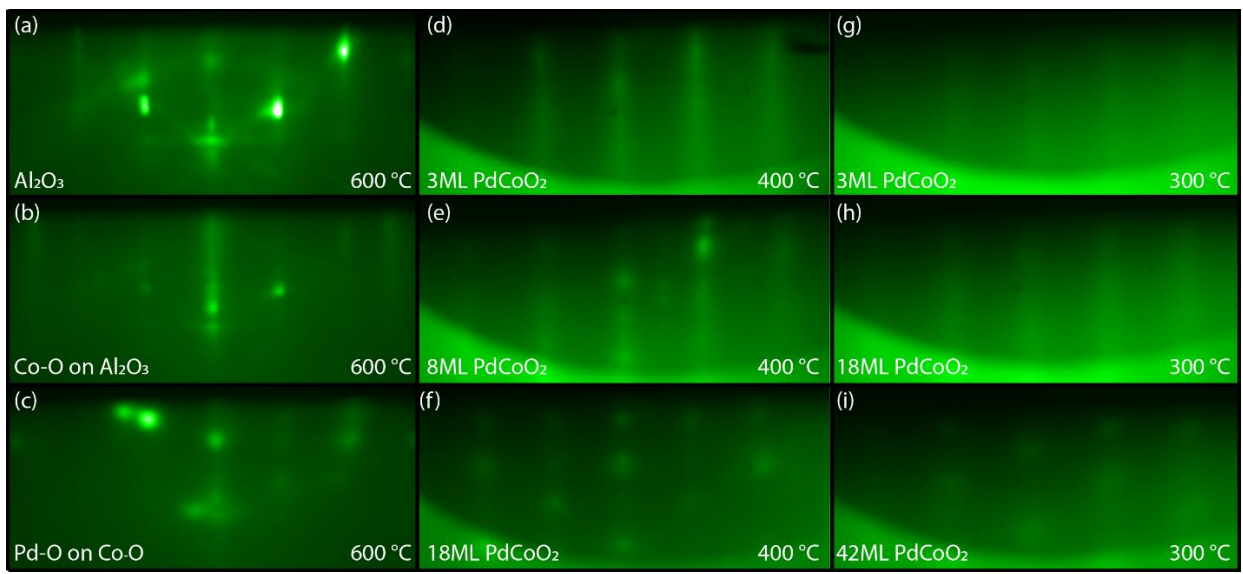

**Fig. 1. Reflection high-energy electron diffraction (RHEED) images for the growth of PdCoO$_2$ on Al$_2$O$_3$ at various conditions.** Images in (a-b) are taken along the Al$_2$O$_3$ <100> and (d-i) are along the PdCoO$_2$ <100> azimuthal direction; temperature, material, and thickness are indicated in the figure. (a) Al$_2$O$_3$ substrate, (b-c) Co-O monolayer (ML) then a Pd-O ML grown at 600 °C; (d-f) 3 ML (d), 8 ML (e), 18 ML (f) PdCoO$_2$ grown at 400 °C; (g-h) 3 ML (g), 18 ML (h), 42 ML (i) PdCoO$_2$ grown at 300 °C.

Figure 2

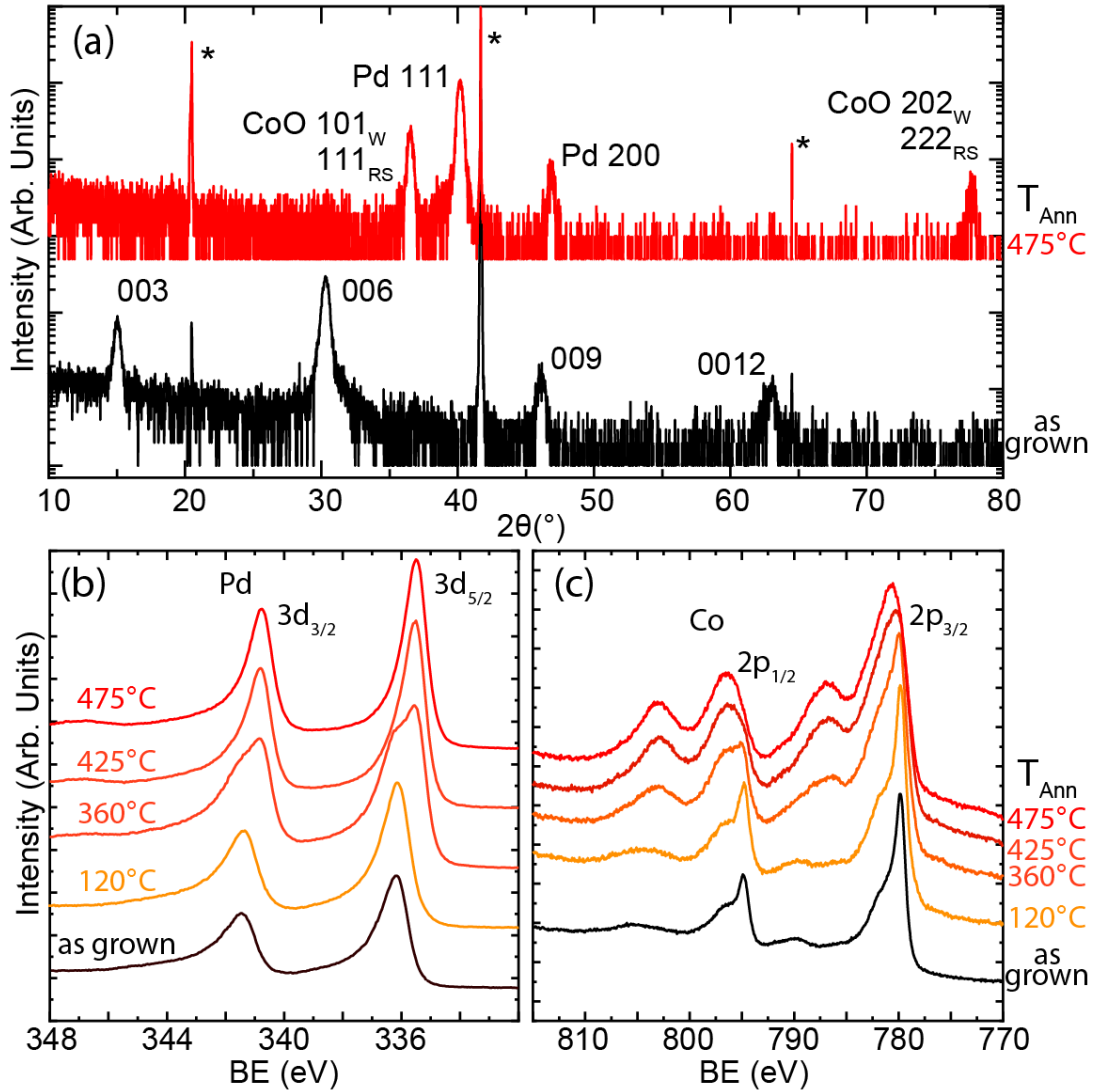

**Fig. 2. Effect of vacuum annealing on the properties of PdCoO$_2$ films.** (a) X-ray diffraction $2\theta$-$\theta$ scans for 25 nm thick PdCoO$_2$ as grown at 300 °C (lower black curve labeled as-grown) and annealed in vacuum to 475 °C (upper red curve). The 003$n$ labels mark the delafossite peaks and the asterisk(*) mark peaks belonging to Al$_2$O$_3$, and the subscripts on CoO indicate wurtzite (W) and rocksalt (RS). Curves are vertically offset for clarity. (b-c) X-ray photoemission spectroscopy about the Pd 3$d$ (b) and Co 2$d$ (c) for the as-grown film and while annealing in vacuum. Curves are vertically offset for clarity.

**Figure 3**

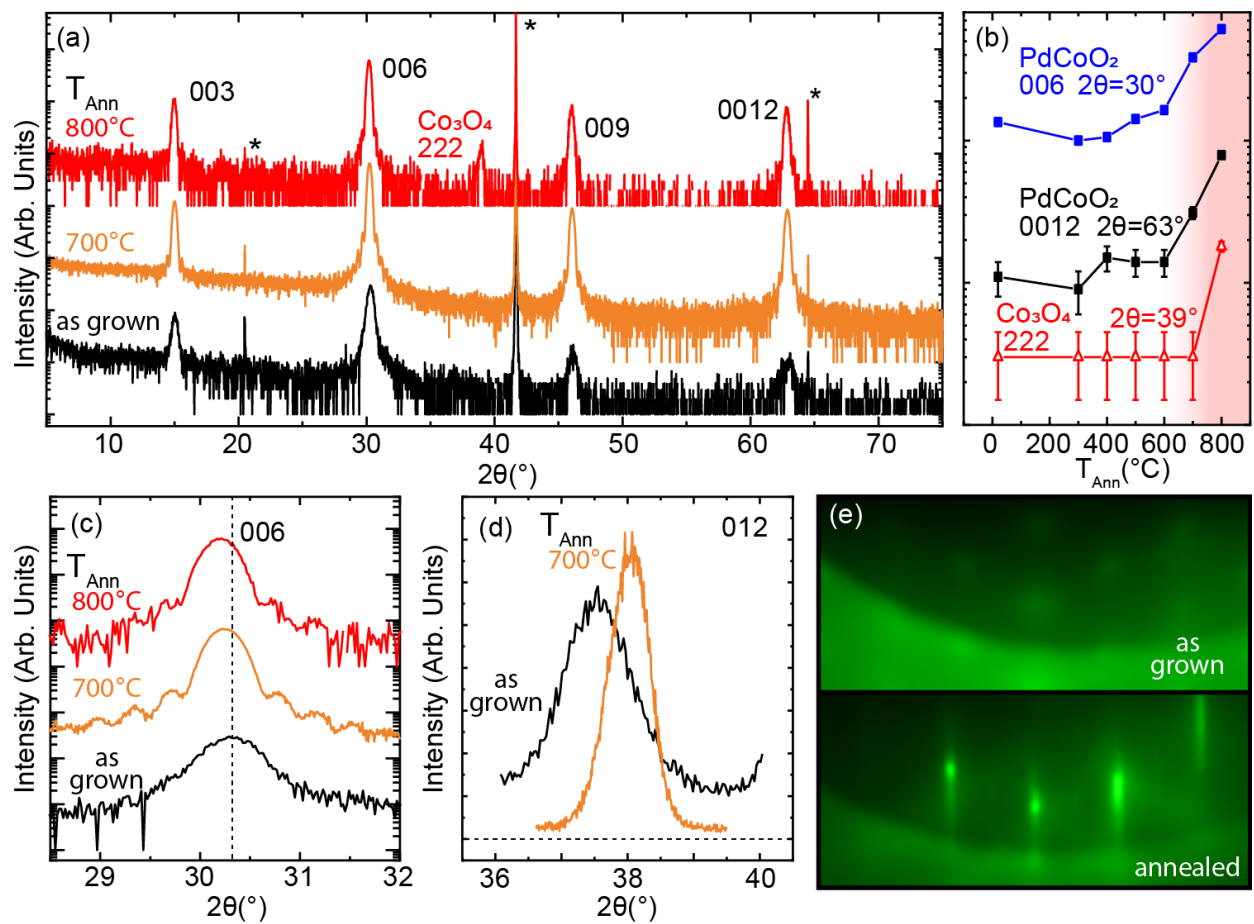

**Fig. 3. Effect of air/oxygen annealing on crystallographic properties for PdCoO₂ films.** (a) X-ray diffraction $2\theta$-$\theta$ scans for 25 nm thick PdCoO$_2$ as grown at 300 °C (lower black curve labeled as-grown), annealed in air to a temperature of $T_{Ann}$ = 700 °C and 800 °C (orange and red curves, respectively). The 003$n$ labels mark the delafossite peaks and the asterisk(*) mark peaks belonging to Al$_2$O$_3$. Curves are vertically offset for clarity. (b) Intensity extracted from the curves in (a) plotted versus annealing temperature $T_{Ann}$ for the 006 and 0012 delafossite and 222 Co$_3$O$_4$ peaks. (c-d) High resolution x-ray diffraction about the 006 (c) and 012 (d) delafossite peaks. (e) Sample RHEED images taken along the PdCoO$_2$ <110> azimuthal direction for a 36 nm-thick film for as-grown (upper panel) versus oxygen annealed (lower panel) (see Supplemental Materials).

**Figure 4**

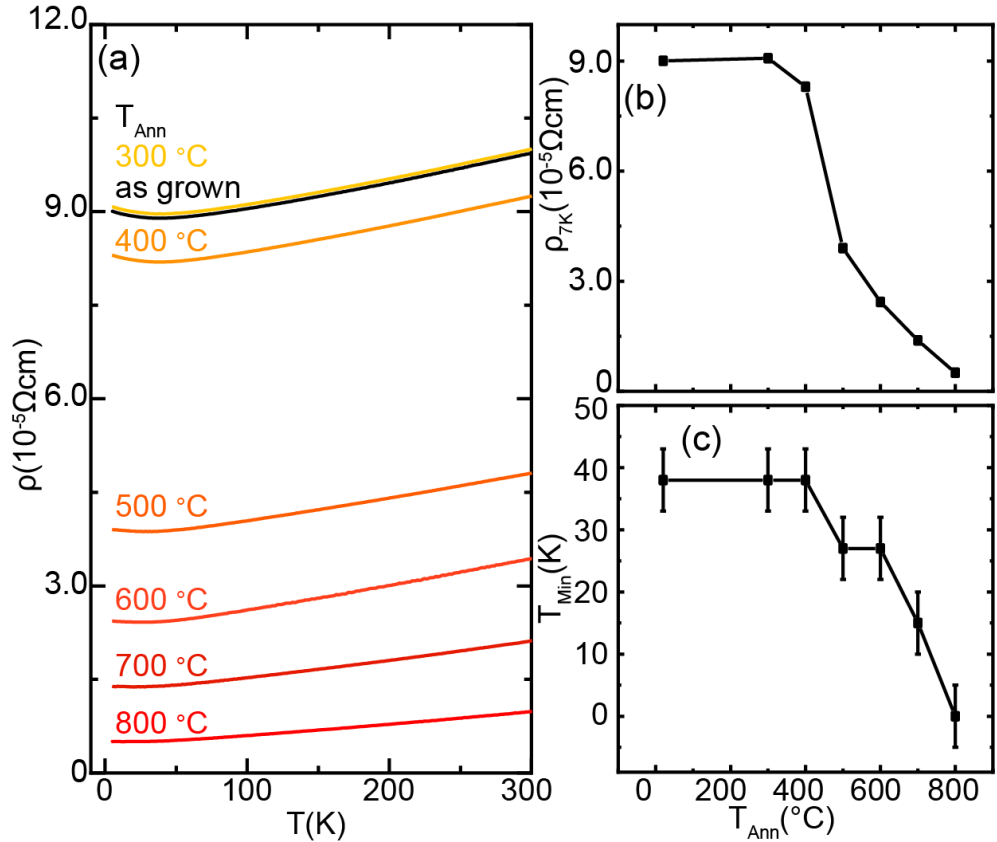

**Fig. 4. Effect of air annealing on the transport properties of a PdCoO$_2$ film.** (a) Resistivity versus temperature data for a 25 nm-thick PdCoO$_2$ film as-grown and annealed in air to a temperature of, $T_{Ann}$ = 300-800 °C. (b) Resistivity as a function of $T_{Ann}$ taken at temperature 7 K. (c) The temperature at which the minimum of resistance occurs (extracted from the data in a) plotted as a function of $T_{Ann}$.

Figure 5

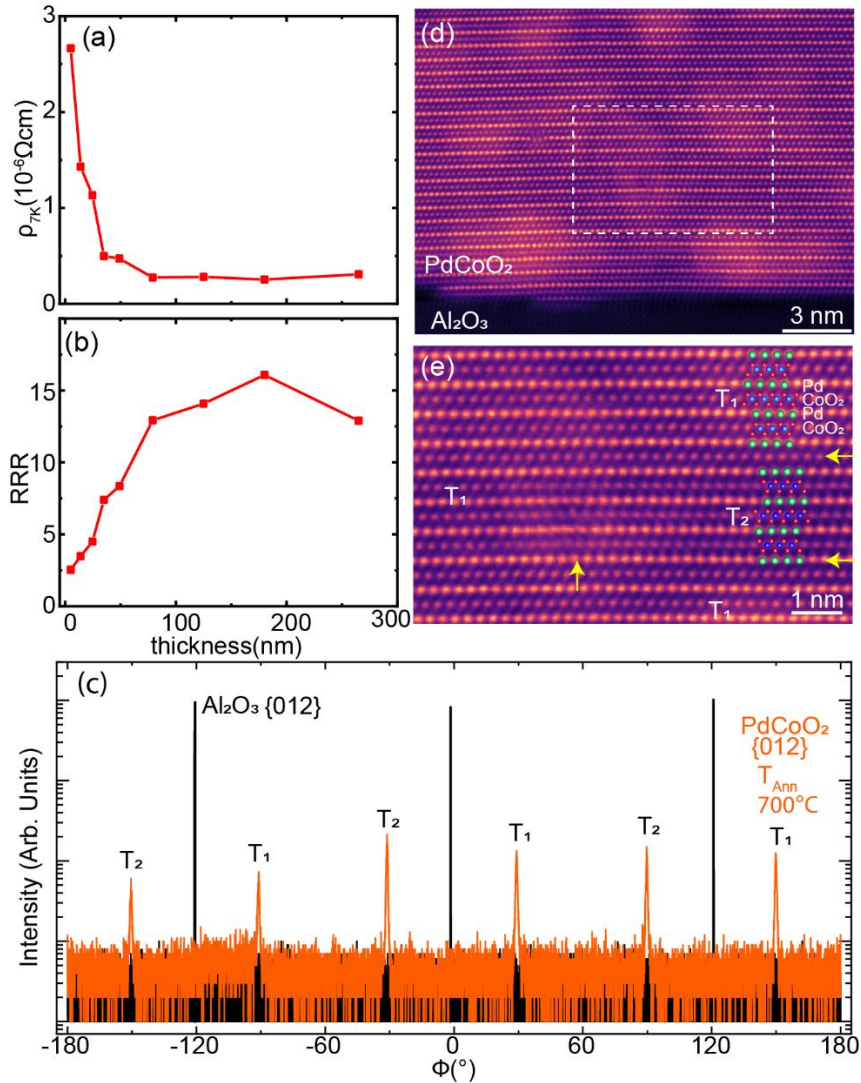

**Fig. 5. Transport and structural properties of air/oxygen annealed PdCoO$_2$ films.** (a-b) Resistivity versus film thickness plots, of PdCoO$_2$ films annealed in oxygen at 800°C, taken at a temperature of 7 K (a) and the residual resistivity ratio (RRR). (c) X-ray diffraction azimuthal scan ($\phi$) taken about the {012} peaks for a PdCoO$_2$ film annealed to 700 °C (orange curve) and the {012} peaks Al$_2$O$_3$ substrate (black curve). The PdCoO$_2$ twin domains are distinguished by the labels $T_1$ and $T_2$. (d-e) HAADF STEM taken along the <100> zone axis. (d) Wide scale image showing the PdCoO$_2$ film and the Al$_2$O$_3$ substrate. (e) Zoom-in of the boxed region in (d). In this image the PdCoO$_2$ atomic positions are overlaid on the image where both $T_1$ and $T_2$ domains are shown, and domain boundaries are highlighted by the yellow arrows.